\documentclass[12pt]{article} 
\begin{document}
\rightline{ UW/PT 96--20 }  
\rightline{September 1996}

\vskip 2cm

\begin{center}
{\bf \large Long Range Physics in a Hot Non-Abelian Plasma}
\end{center}
\medskip
 \begin{center} Patrick~Huet and D.T.~Son \\ {\it Department of Physics \\
  Box 351560 \\ University of Washington\\ Seattle WA 95195}
\end{center}
\bigskip \parbox[]{\textwidth}{ \noindent We derive a set of equations
describing the real time dynamics of modes with spatial momentum of
order $g^2T$ in a high temperature gauge theory, where $g$ is the
coupling constant and $T$ is the temperature.  This dynamics is
stochastic in nature.  Important implications for baryon number
violation at high temperature and for the physics at the electroweak
phase transition, are discussed.  }

\bigskip
\bigskip

\begin{center}{Submitted to: {\it Physics Letters B} }\end{center}

\thispagestyle{empty}
\vfill
\newpage
\setcounter{page}{1}

It has been known for a long time that perturbative calculations in
hot $SU(N)$ gauge theories(QCD) are plagued with an infrared problem.
The scale of momentum where the problem emerges is $g^2T$ where $g$ is
the gauge coupling and $T$ is the temperature of the plasma.  No
reliable way has been suggested for treating these long range modes.
The problem can be seen as originating from the fact that thermal
fluctuations of the magnetic modes at the scale $g^2T$ are large
enough for the theory to become essentially non-linear.  The $g^2T$
scale is important for understanding a large number of physical
problems such as magnetic screening,\footnote{For recent attempts to
compute the magnetic mass, see \cite{nair} and references therein.}
baryon number transitions in the electroweak theory at high
temperature and the determination of the parameters of the electroweak
phase transition. In particular, all of these need to be understood
for a complete calculation of electroweak baryogenesis. The real-time
treatment is appropriate for some problems, especially for computing
topological transition rates in the Standard Model.

In this paper we derive a set of equations which describe the
real-time dynamics of non-perturbative soft modes in hot gauge
theories and we discuss two of their most important applications.  The
first one is baryon number violation at high $T$.  The physics of the
soft modes is stochastic in nature and has the natural time scale
$\sim (g^4 T)^{-1}$.  In particular, our equations suggest that baryon
number violation at high temperature in the unbroken phase scales as
$\alpha_W^5 T^4$ as the result of a diffusive motion across a
topological barrier, rather than $\alpha_W^4 T^4$ as classical
arguments and recent numerical simulations suggest \cite{ak,js}. Last
but not least, our equations constitute a solid computational tool for
the determination of the numerical coefficient of $\alpha_W^5T^4$.
This would have important consequences for electroweak baryogenesis,
and the constraints imposed on it from the structure of the symmetry
breaking sector and from the {\it CP} violating sector.  As a second
application, we introduce a possible path for the real-time resolution
of infrared divergences which plague the perturbative computation of
the parameters of the electroweak phase transition.

\subsection*{Stochastic dynamics of soft modes in hot gauge theory}

{\bf 1.}  Let us begin with a simple argument showing that the time
scale responsible for processes occurring at the space extent of
$(g^2T)^{-1}$ has the natural time scale of $(g^4T)^{-1}$.\footnote{
For parallel considerations with an emphasis on cross barrier
transitions see \cite{asy}.} To this end we consider thermal
fluctuations of the gauge field $A_\mu$.  The power spectrum of these
fluctuations can be related to the imaginary part of the gluon
propagator by the fluctuation--dissipation theorem.  At low
frequencies, $\omega\ll T$, one has,
\begin{eqnarray}
  \langle A_\mu^*(\omega,{\bf q})A_\nu(\omega',{\bf q}')\rangle=
  -{2T\over\omega}\mbox{Im}D^R_{\mu\nu}(\omega, {\bf q})\cdot
  (2\pi)^4\delta(\omega-\omega')\delta({\bf q}-{\bf q}')
  \label{fluc-diss}
\end{eqnarray}
The retarded propagator $D^R$ at small $\omega$ and ${\bf q}$ acquires
important contributions from the hard thermal loops,
\[
  D^R(\omega, {\bf q})\sim{1\over\omega^2-{\bf q}^2-\Pi(\omega,{\bf
q})} \, .
\]
For non-perturbative physics, we are interested only in transverse
(or magnetic) modes, for which the gluon polarization $\Pi(\omega,{\bf
q})$ has the following behavior
\[
  \Pi(\omega,{\bf q})=-icg^2T^2{\omega\over q} \, ,
\]
in the regime $\omega\ll q\equiv|{\bf q}|$, where $c$ is some real
constant.  Eq.\ (\ref{fluc-diss}) implies, in this regime,
\begin{equation}
  \langle A^*(\omega,{\bf q})A(\omega',{\bf q}')\rangle\sim
  {g^2T^3q\over q^6+c^2g^4T^4\omega^2}\cdot
  \delta(\omega-\omega')\delta({\bf q}-{\bf q}') \, .
  \label{fomq}
\end{equation}
To compute the one-time correlation of the field, we integrate
Eq.~(\ref{fomq}) with respect to $\omega$; this gives
\begin{equation}
  \langle A^*(t,{\bf q})A(t,{\bf q}')\rangle\sim
  \delta({\bf q}-{\bf q}')\int\!d\omega\,(-1){T\over\omega}\mbox{Im}
  {1\over\omega^2-q^2+icg^2T^2{\omega\over q}}\sim 
  {T\over q^2}\delta({\bf q}-{\bf q}')
  \label{Atq}
\end{equation}
where the integral is saturated for $\omega\sim g^{-2}T^{-2}q^3$.
Substituting $q\sim g^2T$ in this expression implies $\omega\sim
g^4T$, which is what we wanted to show.  To see that the
non-perturbative scale of spatial momenta is $g^2T$, we can use
Eq.~(\ref{Atq}) to find the contribution of the modes with momenta of
order $q$ to the correlator of $A$ at coinciding points; we find
\[
  \langle A^2(t,{\bf x})\rangle\sim Tq \, .
\]
Hence the typical fluctuation size of $A$ is $(Tq)^{1/2}$ if
restricted to modes with spatial momentum of order $q$.  The nonlinear
contribution to the field tensor $g A^2$ becomes comparable with the
linear part $\partial A$ when $q\sim g^2T$.  So, we have shown that
the non-perturbative physics is associated with modes with spatial
momenta of order $g^2T$ and frequencies of order $g^4T$.

{\bf 2.}  To find the effective theory describing these
non-perturbative modes, we resort to the recent reformulation of the
hard thermal loop Lagrangian in terms of a kinetic Vlasov equation
\cite{bi,kelly}.  In QCD, the soft dynamics is described by the soft field
$A_\mu(t,{\bf x})$ and $\delta n^a(t,{\bf x},{\bf p})$, the deviation
from equilibrium of the hard gluon distribution functions.  The scale
of space and time variation of both $A_\mu(t,{\bf x})$ and $\delta
n^a(t,{\bf x},{\bf p})$ is much larger than $T^{-1}$ while ${\bf
p}\sim T$.  The time evolution of these two effective variables of the
theory is described by a Langevin-type Vlasov equation having the form
\begin{equation}
  v^\mu D^{ab}_\mu\delta n^b(t,{\bf x},{\bf p})+
  g{\bf v}\cdot{\bf E}^a{\partial
  \bar{n}(|{\bf p}|)\over\partial|{\bf p}|}=0
  \label{Vlasov1}
\end{equation}
with
\begin{equation}
  \partial^\mu F^a_{\mu\nu}=2gN\int\!{d{\bf p}\over(2\pi)^3}\,
  v_\nu\delta n^a(t,{\bf x},{\bf p})+\xi^a_\nu (t, {\bf x})
  \label{Vlasov}
\end{equation}
where $v^\mu=(1,{\bf p}/|{\bf p}|)$, and
$D_\mu^{ab}=\delta^{ab}\partial_\mu+gf^{abc}A_\mu^c$ is the covariant
derivative.  In these equations, $\bar{n}$ refers to the
equilibrium Bose-Einstein distribution. The difference between Eq.\
(\ref{Vlasov}) and that discussed in Ref.\ \cite{bi} is the presence
of the stochastic noise $\xi^a_\nu(x)$ in the field equation.  Eqs.\
(\ref{Vlasov1},\ref{Vlasov}) could be derived using the formalism of
Ref.\ \cite{CalzettaHu} which has recently been implemented in the
scalar theory \cite{GleinerMuller}. However, in this letter we will
rather use physical arguments to justify the need to include the
stochastic noise $\xi^a_\nu(x)$ into the kinetic equation and to find
the correlator $\langle\xi^a_\mu(x)\xi^b_\nu(y)\rangle$.  In fact,
while the noiseless Vlasov equation describes the time evolution of
the {\em main values} of the mean field $A_\mu(x)$ and the
distribution function $\delta n^a(t,{\bf x},{\bf p})$, there are many
instances where spontaneous thermal fluctuations need to be considered
(for example, in the problem of baryon number violation in a hot
plasma).  Because the term $\xi^a_\nu(x)$ in Eq.\ (\ref{Vlasov})
reflects fluctuations in the current density, one of our main tasks is
to find the form of the noise correlator and to simplify Eqs.\
(\ref{Vlasov1},\ref{Vlasov}) so that they include only the modes we are
interested in, i.e., modes with $\omega\sim g^4T$ and $p\sim g^2T$.

Before proceeding further, let us make one important remark.  Eqs.\ 
(\ref{Vlasov1},\ref{Vlasov}) have been shown in Ref.\ \cite{bi} to
reproduce the HTL (hard thermal loop) effective Lagrangian describing
the physics of modes with momentum $q\sim gT$.  One could ask
whether the HTL Lagrangian correctly describes the physics at much
smaller spatial momentum scale of $g^2T$.  A diagrammatic analysis
shows that beside the HTL diagrams, diagrams with ladder insertions
are also important (see \cite{ls} and a recent discussion in
\cite{bms}).  However, we argue here that these diagrams do not modify
the dynamics at the momentum scale $g^2T$.  Our first argument is
based on the similarity of the ladder diagrams with those considered
in \cite{Jeon} in the framework of the scalar theory. In the latter
context, it has been shown that the effect of this infinite set of
diagrams is the generation of a collision term in the transport 
equations.  One can argue that this statement can be extended also to 
gauge theories \cite{inpreparation}.  Therefore, we can take into 
account all diagrams relevant to the $g^2T$ dynamics by simply 
adding a collision term to the RHS of Eq.\ (\ref{Vlasov1}).  
However, as will be demonstrated later, the effect of collisions is 
suppressed at length scales $(g^2T)^{-1}$ and starts to becoming 
important only at length scales of order $(g^4T)^{-1}$.  As a bonus, 
we can safely ignore all non-HTL diagrams.  A second argument is based 
on explicit calculations in QED showing that the resumed diagrams do 
coincide with the naive HTL answer for $\Pi(\omega,{\bf q})$ when 
$\omega,{\bf q}\sim g^2T$ \cite{ls}.

{\bf 3.} Let us now derive the correlator of $\xi^a_\nu(x)$.  We begin
by considering in detail a simpler case, namely, hot QED, for which
the Vlasov equation has the form,
\begin{equation}
  \left({\partial\over\partial t}+
  {\bf v}{\partial\over\partial{\bf x}}\right)
  \delta n(t,{\bf x},{\bf p})+e{\bf Ev}
  {\partial\bar{n}(|{\bf p}|)\over\partial|{\bf p}|}=0 
  \label{vlasovqed}
\end{equation}
\begin{equation}
  \partial^\mu F_{\mu\nu}=4e\int\!{d{\bf p}\over(2\pi)^3}
  v_\nu\delta n(t,{\bf x},{\bf p})+\xi_\nu(t,{\bf x}) \, .
  \label{2nd}
\end{equation}
Here $\delta n(t,{\bf x},{\bf p})$ is the deviation from equilibrium
of the distribution function of fermions with a given spin (we assume
that the distribution functions are diagonal in the spin space), and
$\bar{n}$ is the equilibrium Fermi distribution.  The factor 4 is due
to the fact that the plasma consists of particles and anti-particles,
each having 2 spin degrees of freedom.\footnote{The distribution
function of anti-particles deviates from equilibrium by the amount
$-\delta n(t,{\bf x},{\bf p})$.}  The dynamics in QED is simplified,
since the Vlasov equation is linear, but still, it is useful to
consider it before turning to the more complex case of QCD.

To see the physical origin of the stochastic noise $\xi_\nu$ in Eq.\
(\ref{2nd}), let us imagine the plasma as a collection of fermions and
anti-fermions, with, for distribution functions, $n_+(t,{\bf x},{\bf
p})$ and $n_-(t,{\bf x},{\bf p})$, respectively.  In the absence of an
external field, $n_{\pm}(t,{\bf x},{\bf p})$ are equal to the thermal
distribution $\bar{n}(p)$ and the resulting current density in the
plasma vanishes
\begin{equation}
  j_\mu({t,\bf x})=e\int\!{d{\bf p}\over(2\pi)^3}\,v_\mu\sum_\alpha 
  (n^\alpha_+(t,{\bf x},{\bf p})-n^\alpha_-(t,{\bf x},{\bf p})) = 0 \, ,
  \label{current_density}
\end{equation}
(the index $\alpha$ labels the spin degrees of freedom).  However,
Eq.\ (\ref{current_density}) represents only the {\em average} of the
current density: there are always thermal fluctuations of
$j_\mu(t,{\bf x})$ due to fluctuations of the distribution functions
$n_{\pm}(t,{\bf x},{\bf p})$ about the Bose-Einstein main value.
These fluctuations can be divided into a long-distance part and a
short-distance part.  The former is $\delta n(t,{\bf x},{\bf p})$
which enters the Vlasov equation (Eq.\ (\ref{vlasovqed})).  These
fluctuations have for typical length scale $\sim (g^2T)^{-1}$.  The
short-distance part fluctuates over much shorter distance scales (up
to $T^{-1}$).  These short-distance fluctuations are essential for
keeping the soft modes in thermal equilibrium. A Vlasov equation
without the stochastic noise predicts that modes with $\omega<q$ go
away after a sufficiently large amount of time due to Landau damping:
this is not the case of fluctuations in a realistic plasma.  Let us
denote the short-distance fluctuations of the distribution function by
$\Delta n_\alpha(t,{\bf x},{\bf p})$.  Then $\xi_\nu(x)$, in Eq.~
(\ref{2nd}), is the fluctuation of the current density
\[
  \xi_\nu(t,{\bf x})=e\int\!{d{\bf p}\over(2\pi)^3}\,v_\nu\sum_\alpha
  (\Delta n_+^{\alpha}-\Delta n_-^{\alpha}) \, .
\]
One property of $\xi_\nu$ follows from charge conservation,
$\partial_\nu\xi^\nu=0$.  To find the correlation of $\xi_\nu(x)$, we
first observe that the correlation of the thermal fluctuations $\Delta
n_{\pm}(t,{\bf x}, {\bf p})$ at coinciding times is given by
\[
  \langle \Delta n_+^\alpha(t,{\bf x},{\bf p})
  \Delta n_+^{\alpha'}(t,{\bf x}',{\bf p}')\rangle=
  \langle \Delta n_-^\alpha(t,{\bf x},{\bf p})
  \Delta n_-^{\alpha'}(t,{\bf x}',{\bf p}')\rangle=
\]
\begin{equation}
  (2\pi)^3\delta({\bf p}-{\bf p}')\delta({\bf x}-{\bf x}')
  \delta^{\alpha\alpha'}\bar{n}(|{\bf p}|)(1-\bar{n}(|{\bf p}|))
  \label{Delta2}
\end{equation}
(the correlator $\langle\Delta n_+\Delta n_-\rangle$ vanishes).
The spatial delta function $\delta({\bf x}-{\bf x}')$ reflects the
short-distance nature of the fluctuations included in $\Delta
n_\pm(t,{\bf x}, {\bf p})$.  Since we are interested in the correlation
of $\xi_\nu(x)$ at different times, we also have to know the time
evolution of $\Delta n_\pm(t,{\bf x}, {\bf p})$.  It is obtained by
solving the following transport equation\footnote{This equation is
appropriate for describing the propagation of fluctuations on a time
scale much smaller than the transport mean free time.}
\begin{equation}
  \left({\partial\over\partial t}+{\bf v}{\partial\over\partial{\bf x}}
  \right)\Delta n(t,{\bf x},{\bf p}) =0 \, .
  \label{transp_Delta}
\end{equation}
Its solution is $\Delta n(t,{\bf x},{\bf p})= \Delta n(0,{\bf x}-{\bf
v}t,{\bf p})$.  From this we infer that the noise correlator is
\begin{equation}
  \langle\xi_\mu(t,{\bf x})\xi_\nu(t',{\bf y})\rangle
  =4e^2\int\!{d{\bf p}\over(2\pi)^3}\,v_\mu v_\nu
  \delta({\bf x}-{\bf y}-{\bf v}(t-t'))
  \bar{n}(|{\bf p}|)(1-\bar{n}(|{\bf p}|)) \, .
  \label{source_x}
\end{equation}
The chain of arguments which lead to Eq.\ (\ref{source_x}) is somewhat
intuitive.  We can, however, verify this result in various ways.  One 
way is to compute the correlator directly in terms of the fundamental 
fermionic field operator, $\psi$,
\begin{equation}
  \langle\xi_\mu(x)\xi_\nu(y)\rangle=
  e^2\langle\bar{\psi}(x)\gamma_\mu\psi(x)\cdot
  \bar{\psi}(y)\gamma_\nu\psi(y)\rangle
  \label{psibar}
\end{equation}
The RHS of Eq.\
(\ref{psibar}) is easily evaluated in the momentum representation.
It is, in the approximation of free fermions,
\[
  \langle\xi_\mu(\omega,{\bf q})\xi_\nu(\omega',{\bf q}')\rangle
  =4e^2\int\!{d{\bf p}\over(2\pi)^3}\,v_\mu v_\nu
  \delta(\omega-{\bf vq})
  \bar{n}(|{\bf p}|)(1-\bar{n}(|{\bf p}|))
\]
\begin{equation}
  \times (2\pi)^4\delta(\omega-\omega')\delta({\bf q}-{\bf q}')\,.
  \label{source_q}
\end{equation}
It is simple to show that Eq.\ (\ref{source_q}) is just the Fourier
transform of Eq.\ (\ref{source_x}).

We can also verify Eqs.\ (\ref{source_x},\ref{source_q})
using the fluctuation-dissipation theorem.  Schematically, we can solve
Eq.~(\ref{vlasovqed},\ref{2nd}) to relate the power spectrum of the
transverse part of $A_\mu$ with that of $\xi_\mu$ by the very simple
formula,
\begin{equation}
  \langle{\bf A}^*(\omega,{\bf q}){\bf A}(\omega',{\bf q}')\rangle=
  {\langle|\xi(\omega,{\bf q})|^2\rangle\over
  |\omega^2-{\bf p}^2-\Pi(\omega,{\bf q})|^2} \, .
  \label{A2xi2}
\end{equation}
We did check that Eq.~(\ref{A2xi2}) leads to the spectrum of
$A_\mu$--fluctuations satisfying the fluctuation-dissipation theorem,
Eq.\ (\ref{fluc-diss}).

Having obtained the correlator, we now proceed to developing an
effective dynamics for the long range modes of hot QED.  We can solve
the Vlasov equation Eq.~(\ref{vlasovqed}) for $\delta n(t,{\bf x},
{\bf p})$, and use it to obtain an equation for the induced current
responding to the external field $A_\mu$. We obtain
\begin{equation}
  \delta n(t,{\bf x},{\bf p})=-e\int\limits_0^\infty\!du\,
  {\bf vE}(t-u,{\bf x}-u{\bf v})
  {\partial\bar{n}(|{\bf p}|)\over\partial|{\bf p}|} \, ,
  \label{deltanE}
\end{equation} 
and\footnote{Since the important modes are magnetic ones, we can limit
ourselves to the three-vector ${\bf j}$.}
\begin{equation}
  {\bf j}(t,{\bf x})=4e\int\!{d{\bf p}\over(2\pi)^3}\,
  {\bf v}\delta n(t,{\bf x},{\bf p})=
  -{e^2T^2\over12\pi}\int\!d{\bf y}\,
  {{\bf x}-{\bf y}\over|{\bf x}-{\bf y}|^4}[({\bf x}-{\bf y})\cdot
  {\bf E}(t-|{\bf x}-{\bf y}|,{\bf y})] \, .
  \label{jxAy}
\end{equation}

Until now, our only assumption has been that $\omega$ and ${\bf k}$
are much smaller than $T$.  In the regime $\omega\sim g^4T$, ${\bf
k}\sim g^2T$, additional simplification applies.  In Eq.\ (\ref{jxAy})
we can replace $t-|{\bf x}-{\bf y}|$ with $t$, since the typical time
scale is $(g^4T)^{-1}$ and ${\bf x}-{\bf
y}\sim(g^2T)^{-1}\ll(g^4T)^{-1}$.  Substituting (\ref{jxAy}) into Eq.\
(\ref{2nd}) and recalling that $\dot{\bf E}$, $=\ddot{\bf A}$, is
negligible in comparison to ${\bf\nabla}\times{\bf B}$, where ${\bf
B}\equiv{\bf\nabla}\times{\bf A}$ is the magnetic field, we obtain an
equation which evolves long range fields in hot QED,
\begin{equation}
  {\bf\nabla}\times[{\bf\nabla}\times{\bf A}(t,{\bf x})]=
  {e^2T^2\over12\pi}\int\!d{\bf y}\,{ {\bf x}-{\bf y} \over|{\bf
  x}-{\bf y}|^4} [({\bf x}-{\bf y})\cdot\dot{\bf A}(t,{\bf y})]
  +\xi(t,{\bf x}) \, . \label{qed_eq_final}
\end{equation}
Finally, we observe from Eq.~(\ref{source_x}) that the noise
correlator is only non-vanishing on the light cone, $t=\pm|{\bf x}|$.
Since ${\bf x}$ is typically much smaller(by a factor of $g^2$) than
the time scale we are interested in, we can approximate the time
dependence of the noise correlator with a delta function.  To find the
spatial dependence of $\xi_\nu$, we integrate the RHS of Eq.\
(\ref{source_x}) with respect to $t$ and find,
\begin{equation}
  \langle\xi_i(t,{\bf x})\xi_j(t',{\bf y})\rangle=
  {e^2T^3\over 6\pi}{(x-y)_i(x-y)_j\over|{\bf x}-{\bf y}|^4}
  \delta(t-t') \, .
  \label{qed_source_final}
\end{equation}
Eqs.\ (\ref{qed_eq_final},\ref{qed_source_final}) are the final
equations describing the evolution of the soft modes in hot QED.

Before going further, we need to justify the neglect of the collision
term in the Vlasov equation.  The typical time scale for scattering to
significantly affect particle distribution functions, is determined by
the transport cross section and is of order $(g^4T)^{-1}$.  Therefore
one would think that a scattering term should be included in the
Vlasov equation (Eq.\ (\ref{Vlasov})) since this time scale is of the
same order as the time scale under investigation.  However, this is
not so as we now explain.  According to Eq.\ (\ref{jxAy}), the current
density $j_{\mu}(x)$ depends on the background field $A_\mu(y)$ at time
moments $y_0$ so that $x_0-y_0\sim (g^2T)^{-1}$.  This means that, on
the time scale of $(g^4T)^{-1}$, the plasma responds almost
instantaneously to a change in the soft background, and the slow
relaxation due to particle scattering has only a negligible effect.
We can verify this statement by including a simple collision
term to the Vlasov equation,
\[
  \left({\partial\over \partial t}+
  {\bf v}{\partial\over\partial{\bf x}}\right) \delta
  n(t,{\bf x},{\bf p}) +g{\bf Ev}{\partial\bar{n}(p)\over\partial p}= 
  -{1\over\tau}\delta n(t,{\bf x},{\bf p}) \, ,
\]
where $\tau$ is the transport mean free time $\sim (g^4T)^{-1}$, and
observe that its sole effect is the smearing of the delta-function
over $\omega-{\bf vq}\sim g^4T$ in Eq.~(\ref{source_q}).  This has a
negligible effect on the final equation (\ref{qed_eq_final}) and the
noise correlator (\ref{qed_source_final}) (the effect becomes large
only when considering much smaller spatial momentum scale, $g^4T$).
As a result, we can neglect the effect of scattering on the behavior of
the plasma since the scale of spatial momentum responsible for
non-perturbative physics in hot QCD has been shown to be $g^2T$.

{\bf 4.} By now, we have accumulated all the tools to finally address
the dynamics of the soft modes in a non-Abelian plasma in the
non-perturbative regime of interest.  We will consider an SU($N$) pure
gauge theory.  The distribution function of gluons in a plasma is a
matrix $N^{ab}(t,{\bf x},{\bf p})$ in color space\footnote{We assume
that the distribution is trivial with respect to gluon polarization,
see, for example, \cite{Mrowczynski}.}.  The mean value of $N^{ab}$ is
the equilibrium distribution $\bar{n}(|{\bf p}|)\delta^{ab}$, and, as
before, fluctuations of $N^{ab}$ can be divided into soft and hard
components.  The soft component can be written in the form
$-if^{abc}\delta n^c$ where $\delta n$ is the parameter entering the
Vlasov equations (\ref{Vlasov1}, \ref{Vlasov}).  The hard component $\Delta
N^{ab}$ gives rise to the stochastic source in the RHS of
Eq.\ (\ref{Vlasov}),
\begin{equation}
  \xi^a_\mu(t,{\bf x})=ig\int\!{d{\bf p}\over(2\pi)^3}\,v_\mu f^{abc}
  \sum_\alpha\Delta N_\alpha^{bc}(t,{\bf x},{\bf p})
  \label{xiDeltaQCD}
\end{equation}
($\alpha$ denotes gluon polarization).  The one-time correlator of
$\Delta N^{ab}$ is given by a trivial generalization of Eq.\
(\ref{Delta2}),
\begin{equation}
  \langle\Delta N^{ab}_\alpha(t,{\bf x},{\bf p})
  \Delta N^{b'a'}_{\alpha'}(t,{\bf x}',{\bf p}')\rangle=
  (2\pi)^3\delta({\bf p}-{\bf p}')\delta({\bf x}-{\bf x}')
  \delta^{aa'}\delta^{bb'}\delta^{\alpha\alpha'}
  \bar{n}(|{\bf p}|)(1+\bar{n}(|{\bf p|})) \, .
  \label{Delta2QCD}
\end{equation}
Furthermore, $\Delta N^{ab}$ satisfies the transport equation,
\begin{equation}
  v^\mu{\cal D}_\mu\Delta N=0
  \label{transport_Delta}
\end{equation}
where ${\cal D}_\mu=\partial_\mu+[{\cal A}_\mu,\ldots]$, ${\cal
A}_\mu^{ab}=gf^{abc}A^c_\mu$.  The substitution of the covariant
derivative to the space derivative $\partial_x$ in Eq.\
(\ref{transport_Delta}) reflects the fact that hard gluon precesses in
color space when propagating on the soft background.  Remembering that
the field strength $A$ is typically of order $gT$, we observe that
$\partial_x$ is of the same order as $gA\sim g^2T$, in which case, the
$gA$ in the covariant derivative term cannot be dropped.\footnote{In
more physical terms, the color orientation of a hard particle changes
essentially when moving over distances of order $(g^2T)^{-1}$.  In
contrast, the deviation of the particle trajectory from the straight
line on the scale of $(g^2T)^{-1}$ is negligible.  In fact, the
magnetic field $B\sim\partial A\sim g^3T^2$ corresponds to the
curvature radius of the trajectory of hard particles of order
$T/gB\sim(g^4T)^{-1}$ which is much larger than the scale we are
interested in.}  Combining Eqs.\ (\ref{xiDeltaQCD}), (\ref{Delta2QCD})
and (\ref{transport_Delta}), we find for the noise correlator,
\begin{equation}
  \langle\xi^a_\mu(t,{\bf x})\xi^b_\nu(t', {\bf y})\rangle= 2g^2N
  U^{ab}(t,{\bf x}; t',{\bf y})\int\!{ d{\bf p}\over(2 \pi)^3}\, 
  v_\mu v_\nu\delta({\bf x}-{\bf y}-{\bf v} (t-t'))\bar{n}(|{\bf p}|)
  \left(1+ \bar{n}(|{\bf p}|)\right) \, .
  \label{noise_QCD}
\end{equation}
$U(x,y)$ is the Wilson line connecting $x$ and $y$ in the
adjoint representation
\[
  U(x,y)=T\exp\left(-\int\limits_y^x\!dz^\mu\,{\cal A}_\mu(z)\right) \, .
\]
The presence of the Wilson line renders Eq.\ (\ref{noise_QCD}) explicitly
gauge covariant.

The generalization of Eqs.\
(\ref{qed_eq_final},\ref{qed_source_final}) is straightforward.
First, we solve the Vlasov equation Eq.\ (\ref{Vlasov1}) with respect
to $\delta n^a(t,{\bf x},{\bf p})$,
\begin{equation}
  \delta n^a(t,{\bf x},{\bf p}) = 
  -g{\partial\bar{n}(|{\bf p}|)\over\partial|{\bf p}|}
  \int_0\limits^\infty\!du\,U^{ab}(x,x-uv)\,{\bf v}\cdot{\bf E}^b(x-uv)\,.  
\label{solfordelta} 
\end{equation}
We then insert the expression obtained in the expression for the
induced current which appears on the RHS of the field equation
Eq.~(\ref{Vlasov}),
\begin{equation}
  j^a_\nu(t,{\bf x})\, =\, 2gN\int\!{d{\bf p}\over(2\pi)^3}\,
  v_\nu\delta n^a(t,{\bf x},{\bf p})\, . \label{curr}
\end{equation}
The equation obtained is non-local in space and time.  However, we
have already argued that, in the regime $\omega\sim g^4T$, ${\bf k}\sim
g^2T$, the non-locality in time is
confined to a time interval of order $\sim (g^2T)^{-1}$, much smaller
than the typical time scale of variation of the field $\sim
(g^4T)^{-1}$.  Hence, in the limit $g \ll 1$, we can ignore the
non-locality in time.  Inserting Eq.~(\ref{curr}) in the field
equation (Eq.~(\ref{Vlasov})), and using the fact that $\dot{\bf E}$ can
be neglected in comparison to ${\bf D}\times{\bf A}$, we find, after
little algebra,
\begin{equation}
  [{\bf D}\times[{\bf D}\times{\bf A}(t,{\bf x})]]^a=
  {g^2T^2N\over12\pi}\int\!d{\bf y}\, {{\bf x}-{\bf 
  y}\over|{\bf x}-{\bf y}|^4}\,U^{ab}(t,{\bf x},{\bf y})\dot{\bf 
  A}^b(t,{\bf y})\cdot({\bf x}-{\bf y})+\xi^a(t,{\bf x}) 
  \label{final_eq}
\end{equation}
with the noise
correlator (\ref{noise_QCD}) taking the form,
\begin{equation}
  \langle\xi^a_i(t,{\bf x})\xi^b_j(t',{\bf y})\rangle=
  {g^2T^3N\over 6\pi}U^{ab}(t,{\bf x},{\bf y})
  {(x-y)_i(x-y)_j\over|{\bf x}-{\bf y}|^4}\delta(t-t') \, .
  \label{final_source}
\end{equation}
It can be checked that Eq.\ (\ref{final_source}) is consistent with
the transversality condition $D_i\xi_i=0$.\footnote{This condition is
required for Eq.\ (\ref{final_eq}) to have solutions with respect to
$\dot{\bf A}$.}  Equations (\ref{final_eq}) and (\ref{final_source}),
constitute the highlight of our paper.

{\bf 5.}  In summary, we have derived a Langevin-type equation, Eq.\
(\ref{final_eq}), which describes the non-perturbative behavior of
very soft modes in hot gauge theories.  They are non-local in space,
but local in time.  Equation (\ref{final_eq}) is a first-order
differential equation with respect to $t$, and $\dot{\bf A}$ appears
in the equation in a linear fashion (there is no terms like $\dot{\bf
A}^2$).  The noise correlator is local(white) in time, but has a
non-trivial spatial dependence.  Eq.\ (\ref{final_eq}) can be used for
simulating the dynamics of soft modes in a variety of contexts.  To do
so, one generates, at any given time, a noise $\xi$, transverse and
distributed according to Eq.\ (\ref{final_source}), then runs the
system to the next time-slice using Eq.\ (\ref{final_eq}); one repeats
these steps until the system reaches a stationary state.  The
benefit of our formalism is that one can concentrate on relevant modes
of spatial momenta of order $g^2T$ only, since all higher modes have
already been integrated out in deriving Eqs.\
(\ref{final_eq},\ref{final_source}).  Therefore, one can use a lattice
with the lattice cutoff $a^{-1}$ of order $g^2T$, but not $T$ or $gT$
as in most of the methods suggested so far (the hard modes have been
taken into account analytically), which leads to the possibility of
doing calculations with a much larger lattice spacing.  However, to
actually use Eq.\ (\ref{final_eq}) for simulating the dynamics of soft
modes, one has to invert the linear operator acting on $\dot{\bf A}$
on the RHS of Eq.\ (\ref{final_eq}).  This seems to be a potentially
non-trivial problem for numerical calculations, and we hope that this
obstacle can be overcome.

\section*{Application I: Baryon number violation at high $T$}

{\bf 1.} Our main motivation for developing an effective theory for
soft modes is their involvement in numerous important phenomena taking
place in a high temperature environment, one of which being the
mechanism of baryon number violation at high $T$. Existing numerical
simulations, and in particular the most recent ones \cite{ak,js},
suggest that, $\Gamma_w$, the rate of baryon number violation per
unit volume, is $\kappa \, \alpha_w^4 T^4$ with $\kappa \sim 1$. This
corroborates the most naive dimensional estimates which suggest that
$\Gamma$ scales as $\sim \xi^4$ where $\xi$,  $\sim (g^2T)^{-1}$, is a
natural length scale of the problem.  Two reasons suggest caution:
\begin{enumerate}
\item[{-- 1 --}] 
First, the above naive dimensional analysis does not
recognize that the relevant time scale is $(g^4 T)^{-1}$ rather than
$(g^2T)^{-1}$.
\item[{-- 2 --}] Second, existing numerical simulations incorporate high 
momenta modes classically.  This method is potentially sensitive to an 
ultraviolet cutoff and needs to be implemented with great care in 
order not to induce lattice artifacts\cite{bms}.
\end{enumerate}

The effective theory developed in the previous section, has the virtue 
of incorporating the physics of the hard modes into the physics of the 
soft modes, hence, it alleviates the sensitivity on very short distance 
fluctuations.  The outcome is a Langevin-type equation (one derivative 
in time, two derivatives in space) which suggests a cross-barrier 
transition of a diffusive type with natural length scale and time 
scale of order $(g^2T)^{-1}$ and $(g^4T)^{-1}$, respectively.

We are confident that our results will become a tool to further study
the rate of baryon number violation at high $T$.  The qualitative and
quantitative understanding of the latter has important consequences,
some of which are illustrated below.

{\bf 2.} Let us consider the outset of electroweak baryogenesis.  The
most recent computation of the baryon asymmetry produced at the
electroweak era, in the framework of some supersymmetric
theories\cite{huetnelson}, concluded that in a region of parameter
space where all squarks are heavy and nearly degenerate ($ > 100$ GeV),
the baryon asymmetry scales as $\sin \theta_{cp}\, \Gamma_w/\Gamma_s$.
Here, $\Gamma_s$ is the rate per unit volume of violation of the axial
charge due to strong anomalous $QCD$ processes and $\theta_{cp}$ is a
{\it CP} violating phase. In contrast, in a regime of non-degeneracy,
with, say, the super-partners of the third generation of quarks lighter
than $T$ and lighter than the other squarks, the baryon asymmetry
scales as $\sin \theta_{cp} \, \Gamma_w $.

In the {\it speculative} situation where $\Gamma_{w,s}$ scales as
$\alpha_{w,s}^5$ rather than as $\alpha_{w,s}^4$, the result of these
computations is offset in respect to those quoted in Ref.\
\cite{huetnelson}, by a factor of $\alpha_w/\alpha_s \sim 1/3$ in the
degenerate case and a factor of $\alpha_w \sim 1/30$, in the
non-degenerate case. In the latter case, these corrections are very
significant as they correspondingly shift the quoted lower bounds on
the {\it CP} violating phase $\theta_{cp}$ from a range
$10^{-4}-10^{-2}$ to a range $\sim 30$ larger, bringing these bounds
to a ``dangerous'' overlap with the quoted lower bounds obtained from
electric dipole moment experiments\cite{dipole}.

{\bf 3.} Finally, let us briefly speculate on another possible
interesting consequence for electroweak baryogenesis. Another source
of significant constraint on electroweak baryogenesis is the
preservation of the baryon asymmetry at times subsequent to its
production. This requires that baryon number violating processes be
suppressed in the broken phase after the transition. Currently, this
requirement yields very strong constraints on the parameters of the
Higgs sector. It remains to be explored whether the above
considerations will affect the determination of these constraints.

\section*{Application II: Parameters of the electroweak phase transition}

Let us now turn to a second important application of our formulation
of long range physics in a non-Abelian plasma: the determination of
the parameters of the electroweak phase transition. We refer, more
specifically, to the order of the transition, and, if first order, to
the nucleation rate, and the dynamical properties of the phase
interface such as final Higgs expectation value, shape, profile and
velocity. The knowledge of these quantities is required for a
quantitative study of electroweak baryogenesis.  One of the
computational tools often used in their calculation is the equilibrium
free energy, $V(\phi,T)$, of the system.  The perturbative
determination of the latter is known to be plagued with infrared
divergences.  Those arising from the physics at scale $gT$ can be
accounted for perturbatively by resummation of hard thermal
loops\cite{carr,dhlll}.  Others, arising at scale $g^2T$ in the
magnetic components of the gauge sector, remain unaccounted for
perturbatively; their presence is an obstacle to a complete
understanding of the outcome of the electroweak phase transition.

The stochastic dynamics of soft modes presented above sheds
new light on these issues and provides a new tool for resolving
them. To illustrate how, we begin by briefly presenting a real-time
calculation of the free energy $V(\phi,T)$.

{\bf 1.} We consider a gas of gauge bosons with mass $m = g\phi/2$, in the 
presence of a space-dependent Higgs background $\phi(x)$ which 
interpolates smoothly between phases of broken and unbroken 
electroweak symmetry: $\phi(-\infty) = 0$, $\phi(+\infty)= \phi_0$ and 
$\partial_x \phi(\pm\infty)=0$.  Focusing on the gauge components of 
the plasma and their interactions with the kink, we can simultaneously 
describe the dynamics of the kink and of the particles distributions 
$n^a(\bf{x},{\bf p})$ with the following equations\cite{lmt}

\begin{eqnarray}
  {\bf v}{d\over d{\bf x}}n^a({\bf x},{\bf p}) -{1 \over 2
  \varepsilon_{\bf p}} {\bf v}{dm^2(x) \over d{\bf x}}{d\over
  d\varepsilon_{\bf p}}n^a(\bf{x},{\bf p})= C(\tau) \\ -
  \partial^2_{\bf x} \phi(x)= -{\partial V(\phi) \over \partial \phi}
  + {dm^2(x) \over dx}\, \int\!{d{\bf p}\over(2\pi)^3} {1 \over
  \varepsilon_{\bf p}} \, n^a(\bf{x},{\bf p}) \, ,
\label{vlasmotion}
\end{eqnarray}
where $\varepsilon_{\bf p}$ is the space-dependent energy of a
particle with momentum ${\bf p}$ at point $\bf{x}$. The force term in
the first Vlasov equation above is derived from energy conservation,
which, in the kink frame, implies ${\bf p^2}+m^2(x) = $ constant. 
We assume the existence of a stationary situation. This is guaranteed
dynamically by the presence of the collision term $C(\tau)$. Without
the latter, the kink would continuously accelerate leading to an
unpleasant situation\cite{lmt,dhlll}.

The above equations supplemented with appropriate boundary conditions
for the kink and with $\varepsilon_{\bf p}(-\infty) = {\bf p^2}$, not
only provide a unique solution to the propagation of a phase interface
(in a regime where the curvature can be ignored) such as its velocity
relative to the plasma\footnote{Eqs.~(\ref{vlasmotion}) were used in
Ref.~\cite{lmt} to compute the relative motion of the kink in the
plasma.} and its profile, but also provide us with a powerful scheme
for computing $V(\phi,T)$.  A useful quantity to look at, is the space
variation of $T^{\mu\nu}(x)$, the stress energy tensor of the
system. The leading order term of its expansion in $v_k$, the relative
kink-plasma velocity, yields the free energy of the system $
T^{xx}(x)-T^{xx}(-\infty) = V(0,T) - V(\phi(x),T) + v_k L(\phi(x),T) +
{\cal O}(v_k^2) $.  Because the ``drag force'' $L(\phi(x),T)$ vanishes
as the collision term in the first Vlasov equation vanishes, the
effective potential is the solution of Eq.~(\ref{vlasmotion}) in the
collisionless limit.  The solution is $\varepsilon_{\bf p}^2(x) = {\bf
p^2} + m^2(x)$ and, $V(\phi,T)$ given by the so-called ``one-loop''
effective potential\cite{lmt}.  Its minimum, $\phi_0$, scales as
$\phi_0 \sim gT\ m_H^2/m_W^2 $ which, in turn, fixes the size of the
kink to be of the order of $L_k \sim \int d\phi/\sqrt{V(\phi,T)} \sim
1/g^2T \ (m_H/m_W) $.  This approximation of the free energy has been
known for a long time to be inadequate as it does not incorporate any
of the long range plasma physics.  In particular, in the limit $\phi
\to 0$, there is an uncontrollable number of soft quanta $\sim
T/g\phi$ whose interactions (with strength $g^2$) among themselves and
with the hard modes yield uncontrollably large corrections to the free
energy.  The remaining of this section addresses that question.

{\bf 2.} {\it Inclusion of the $gT$--scale physics}. Plasma
physics at the scale $\sim gT$ (``plasmon physics'') becomes relevant when
$g \phi \sim gT$, that is for $\phi < T$. From our estimate of $\phi_0$
above, we see that this physics affects the determination of the free
energy for all values of $\phi \leq \phi_0$, i.e., across the whole
kink. The resolution of these difficulties is
know\cite{carr,dhlll}. One has to incorporate hard thermal loops in
the calculation of $V(\phi,T)$. In the above real time framework, this
operation can be performed very simply by noting that plasmon physics
is contained within a distance $\sim 1/gT \, < \, L_k \sim 1/g^2T$:
the plasmon physics is ``local'' inside the kink and is in a large
extent independent on the properties of the latter.  One can then
describe the kink as coupled to a gas of free bosons with longitudinal
modes replaced with plasmons.  The dispersion relation of a plasmon
is, to a sufficient approximation: $\varepsilon_{\bf p}^2 = {\bf p^2}
+ \Pi ({\bf p}=0)$, where $\sqrt{\Pi ({\bf p}=0)}$ is the plasmon
frequency, $\sim gT$.  Incorporating this modified dispersion relation
in the boundary conditions at $ x \to -\infty$, yields an expression
for the source $s(x)$, which, after simple algebra, reproduces a form
of the free energy known in the literature as the ``one loop
improved'' effective potential\cite{dhlll}. This is a significant
improvement because gauge interactions among plasmons are computable
perturbatively, their effective strength scales according to $g^2 \
T/\sqrt{\Pi ({\bf p}=0)}\sim g <1$. There remains an infrared behavior
in the transverse components of the gauge sector.

{\bf 3.} {\it Inclusion of the $g^2T$--scale physics}. It is
transparent from the above analysis that the calculation of
$V(\phi,T)$ involves the natural scale\footnote{ This scale is not an
artifact of our scheme of computation, not only does this scale set
the size of the interface between two phases but it also fixes the
size of the critical radius in nucleation theory.}  $L_k \sim 1/g^2T$
which is of the order of the one characteristic of the non-linearity
in a hot gauge plasma. Any determination of $V(\phi,T)$ which claims
to incorporate the latter physics, is doomed to be intertwined with
the physics of the $g^2T$ modes, in contrast with the situation we
encountered with the $gT$ modes.  A path for a resolution is provided
by evolving simultaneously the plasma in a Higgs background and in a
soft gauge background. The equations consist in a generalization of
Eqs. (\ref{final_eq}) and (\ref{final_source}) to the presence of a
space-dependent background Higgs field, that is, a Vlasov equation
supplemented with the gauge field equations and the Higgs equation of
motion. In contrast with the situation with the $gT$--modes, these
last two equations do not decouple from each other because the typical
lengths of variation of both backgrounds are comparable, $\sim
(g^2T)^{-1}$. In particular, the space derivative on the LHS of Eq.\
(\ref{vlasmotion}) is to be substituted with a covariant
derivative. These equations, whose study is beyond the scope of this
letter, constitute a natural real-time framework to incorporate the
soft gauge physics in the calculation of the thermodynamics of the
electroweak phase transition.

{\bf Acknowledgments} P.H. would like to acknowledge Raju Venugopalan
for useful conversations.  D.T.S. thanks P.~Arnold and L.~Yaffe for
stimulating discussions.  Work supported by the Department of Energy,
contract DE-FG03-96ER40956.

\end{document}